\documentclass[sigconf,screen,nonacm,review=false,timestamp=false]{acmart}
\AtBeginDocument{%
  }

\setcopyright{acmlicensed}
\copyrightyear{2025}
\acmYear{2025}
\acmDOI{XXXXXXX.XXXXXXX}
\acmConference[ISWC '25]{Make sure to enter the correct
  conference title from your rights confirmation email}{October 14--16,
  2025}{Espoo, Finland}
\acmISBN{978-1-4503-XXXX-X/2018/06}

\usepackage{balance}

\begin{document}

\title{Heatables: Effects of Infrared-LED-Induced Ear Heating on Thermal Perception, Comfort, and Cognitive Performance}

\author{Valeria Zitz}
\email{valeria.zitz@kit.edu}
\orcid{0009-0004-1158-861X}
\affiliation{%
  \institution{Karlsruhe Institute of Technology}
  \city{Karlsruhe}
  \country{Germany}
}

\author{Michael Küttner}
\email{michael.kuettner@kit.edu}
\orcid{0009-0000-9021-0359}
\affiliation{%
  \institution{Karlsruhe Institute of Technology}
  \city{Karlsruhe}
  \country{Germany}
}

\author{Jonas Hummel}
\email{jonas.hummel@kit.edu}
\orcid{0009-0005-8563-6175}
\affiliation{%
  \institution{Karlsruhe Institute of Technology}
  \city{Karlsruhe}
  \country{Germany}
}

\author{Michael T. Knierim}
\email{michael.knierim@kit.edu}
\orcid{0000-0001-7148-5138}
\affiliation{%
  \institution{Karlsruhe Institute of Technology}
  \city{Karlsruhe}
  \country{Germany}
}

\author{Michael Beigl}
\email{michael.beigl@kit.edu}
\orcid{0000-0001-5009-2327}
\affiliation{%
  \institution{Karlsruhe Institute of Technology}
  \city{Karlsruhe}
  \country{Germany}
}

\author{Tobias Röddiger}
\email{tobias.roeddiger@kit.edu}
\orcid{0000-0002-4718-9280}
\affiliation{%
  \institution{Karlsruhe Institute of Technology}
  \city{Karlsruhe}
  \country{Germany}
}

\renewcommand{\shortauthors}{Zitz et al.}

\begin{abstract}
  Maintaining thermal comfort in shared indoor environments remains challenging, as centralized HVAC systems are slow to adapt and standardized to group norms. Cold exposure not only reduces subjective comfort but can impair cognitive performance, particularly under moderate to severe cold stress. Personal Comfort Systems (PCS) have shown promise by providing localized heating, yet many designs target distal body parts with low thermosensitivity and often lack portability. In this work, we investigate whether targeted thermal stimulation using in-ear worn devices can manipulate thermal perception and enhance thermal comfort. We present \textit{Heatables}, a novel in-ear wearable that emits Near-Infrared (NIR) and Infrared (IR) radiation via integrated LEDs to deliver localized optical heating. This approach leverages NIR-IR’s ability to penetrate deeper tissues, offering advantages over traditional resistive heating limited to surface warming. In a placebo-controlled study with 24 participants, each exposed for 150 minutes in a cool office environment (\( \approx 17.5\,^{\circ}\mathrm{C} \)) to simulate sustained cold stress during typical sedentary office activities, Heatables significantly increased the perceived ambient temperature by around \(1.5\,^{\circ}\mathrm{C} \) and delayed cold discomfort. Importantly, thermal benefits extended beyond the ear region, improving both whole-body comfort and thermal acceptability. These findings position in-ear NIR-IR-LED-based stimulation as a promising modality for unobtrusive thermal comfort enhancement in everyday contexts.

\end{abstract}



\begin{teaserfigure}
  \includegraphics[width=\textwidth]{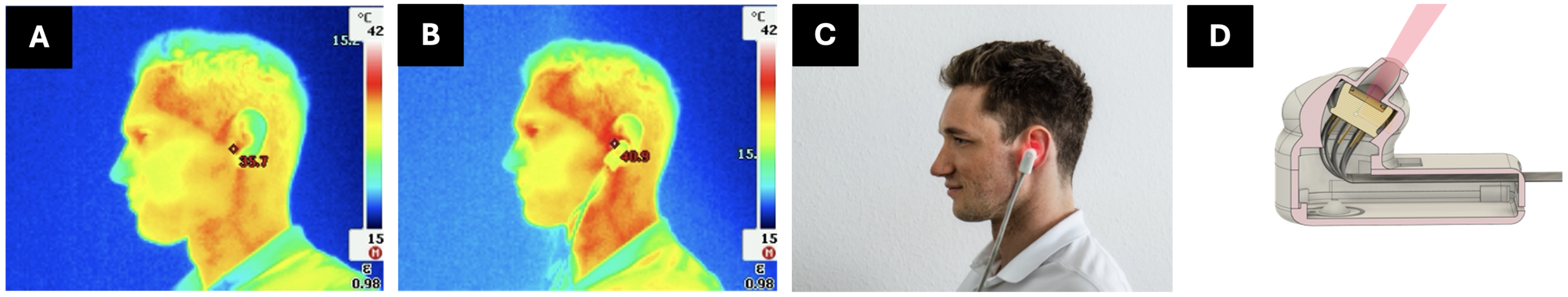}
  \caption{Visual overview of experimental conditions and prototype design, captured using a Testo 883-1 infrared thermal camera. (A) Infrared image of a participant wearing the placebo device after 45 minutes of cold exposure. (B) Infrared image of a participant wearing the active Heatables device after 45 minutes of cold exposure, highlighting localized warming around the ear region. (C) Photograph of a participant wearing the Heatables device. (D) Cross-sectional schematic of the Heatables prototype.}
  \Description{Composite figure showing (A) thermal image with placebo, (B) thermal image with Heatables, (C) photograph of participant wearing active Heatables, and (D) schematic cross-section of the device prototype.}
  \label{fig:teaser}
\end{teaserfigure}

\received{20 February 2007}
\received[revised]{12 March 2009}
\received[accepted]{5 June 2009}

\maketitle

\section{Introduction}
In shared indoor spaces, Heating, Ventilation, and Air Conditioning (HVAC) systems adapt slowly and are typically optimized for group norms, often leading to discomfort and thermal conflicts in multi-user settings~\cite{ashrae1992, iso7730, von_frankenberg_towards_2021}. Amid rising energy costs, many households reduce heating, relying on partially warmed spaces---conditions that heighten thermal vulnerability~\cite{VONPLATTEN2025104013}. Cold exposure not only impairs comfort but also cognitive performance, including memory, attention, and reaction time, with some effects persisting post-rewarming~\cite{donnan_effects_2021, jones_cold_2017, sun_human_2022}.

\begin{figure*}[!t]
  \centering
  \includegraphics[width=\linewidth]{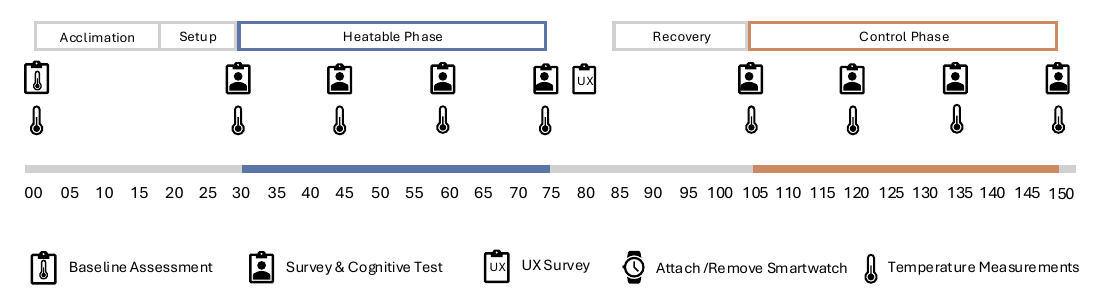}
  \caption{Study procedure illustrating the mixed factorial design. Each participant completed one stimulation session (either Heatables or Placebo) and one control session (no device), with the order of exposure counterbalanced across participants. Measurements were taken at four standardized time points (0, 15, 30, and 45 minutes) during each session.}
  \Description{Diagram of the experimental timeline showing counterbalanced conditions and scheduled measurement intervals.}
  \label{fig:study-procedure}
\end{figure*}

These challenges underscore the need for energy-efficient Personal Comfort Systems (PCS) that deliver localized heating without conditioning entire rooms~\cite{zhang_review_2015, newby_body_2025}. Yet many PCS remain bulky or target thermally insensitive distal regions, limiting daily usability.
Wearables like headphones and earbuds offer new opportunities for seamless thermal stimulation~\cite{roddiger_sensing_2022, schwambach_acceptance_2022}. The auditory canal is richly vascularized and located near the hypothalamus, enabling efficient heat transfer and thermoregulatory coupling~\cite{mu_applied_2021, avelar2013}. Combined with the tissue penetration of near- and infrared (NIR/IR) radiation~\cite{grandinetti_thermal_2015, ito_assessment_2000}, the ear becomes a promising site for perceptual thermal modulation.

We propose a new approach: delivering deep-tissue heat through in-ear wearables. \textit{Heatables} use IR/NIR-LEDs to optically warm the auditory canal, enabling deeper, more distributed stimulation than resistive heating~\cite{grandinetti_thermal_2015}. In a placebo-controlled study under cool office conditions (\(\approx 17.5\,^\circ\mathrm{C}\), 150 minutes), we tested whether (1) in-ear IR/NIR stimulation elevates subjective temperature estimates (ear, wrist, ambient), (2) improves thermal perception, acceptance, and comfort, and (3) preserves cognitive performance, assessed via the Stroop Color and Word Test.

\section{Related Work}

\paragraph{Personal Comfort Systems}
Personal Comfort Systems (PCS) provide individualized localized heating or cooling tailored to individual needs, enabling improved comfort and reduced energy consumption. Examples include heated office chairs \cite{zhang_review_2015, he_heating_2018}, thermoelectric clothing \cite{newby_body_2025, li_performance_2023}, and wearable leg heaters \cite{wang_experimental_2022}. While effective in raising localized temperatures, many PCS target body regions such as wrists or legs that exhibit relatively low thermal sensitivity.

\paragraph{Thermal Stimulation via Head and Ears}
Recent Human-Computer Interaction (HCI) research has identified the head and ears as highly thermosensitive targets. Studies have explored periauricular feedback to enhance comfort or modulate affect
\cite{akiyama_thermon_2013, nasser_thermearhook_2021, stanke_can_2023}. Knierim et al. \cite{knierim_warmth_2024} demonstrated that over-ear thermal stimulation via headphones can enhance both local and whole-body thermal comfort during desk work, comparing active heating (via embedded elements) to passive insulation.

However, these approaches primarily target surface warming around the outer ear, not the vascularized tissues deeper within the auditory canal. While the auditory canal is compact, it contains richly perfused epithelial and cartilaginous tissue and is thermally coupled to core regulation via the tympanic membrane and surrounding vasculature \cite{avelar2013, mu_applied_2021}. These anatomical features make it a promising site for thermal stimulation, particularly with NIR-IR radiation that can penetrate several millimeters into soft tissue \cite{tanaka_beneficial_2013}. In-ear thermal stimulation thus remains largely underexplored, despite its potential to influence thermal perception through both local deep-tissue heating and anatomical proximity to thermoregulatory centers.


\paragraph{Optical Heating for Deep Tissue Warming}
Near-Infrared (NIR) and Infrared (IR) light are well-established for their ability to penetrate biological tissues \cite{grandinetti_thermal_2015, ito_assessment_2000}. Therapeutic applications of NIR-IR radiation—such as pain relief and improved circulation—highlight its potential for safe and effective thermal stimulation \cite{tanaka_beneficial_2013}.

\section{Heatables}
We introduce Heatables, custom in-ear wearables delivering localized optical heat
(see \autoref{fig:teaser}). The system resembles a conventional pair of wired earphones and builds on the in-ear form factor of OpenEarable 2.0 \cite{roddiger_openearable_2025}. To accommodate individual ear anatomies, maximize stability, and improve fit and comfort Heatables can be fitted with ear tips of varying sizes, and optional wingtips.

Each earpiece integrates a multichip LED (MTMD6894T38\footnote{\url{https://www.mouser.de/datasheet/2/1094/MTMD6894T38-2255938.pdf}}) emitting at 670 nm (red), 810 nm (IR), and 950 nm (NIR). For stimulation, only the 810 nm and 950 nm channels were activated and modulated via PWM at 100 Hz (10 ms cycle, 59\% duty cycle) with a forward current of 300 mA. This yielded an average input power of ~480 mW per earpiece. Given the LEDs’ limited optical efficiency (~28\% at 810 nm, ~9\% at 950 nm), it should be noted that the conversion to optical power of the LEDs is inherently limited. The system is USB-powered and includes an Arduino UNO, custom PCBs, thermal safeguards, and interchangeable silicone tips for anatomical fit.


\section{Evaluation}

To investigate the perceptual and cognitive effects of in-ear thermal stimulation, we conducted a mixed factorial study with one between-subjects factor (stimulation condition: Heatables vs. Placebo) and one within-subjects factor (device exposure: Device vs. No Device). Participants were randomly assigned to four counterbalanced experimental groups (stimulation × device) to control for order effects and perceptual expectancy biases.

\subsection{Experimental Setting}

Experiments were conducted in a thermally controlled office environment maintained at approximately 17.5\textdegree C. This ambient temperature is substantially below typical operative temperature ranges observed in European office settings, where comfort temperatures generally range between 22.5\textdegree C and 25\textdegree C during occupancy \cite{Kalz2014}.

Participants wore light to moderately insulating office clothing (e.g., T-shirt or pullover and jeans) and engaged in sedentary desk-based activities such as typing and reading, corresponding to a low physical exertion level of approximately 1.0–1.2 Metabolic Equivalents of Task (MET) \cite{liu_energy_2022,ainsworth_2011}. Given this low metabolic heat production, the environment constituted a sub-neutral thermal context likely to induce gradual cold discomfort over time.

Participants were seated individually at standardized workstations, each equipped with a laptop and external input devices, ensuring consistent posture and activity profiles across all sessions.

To ensure perceptually meaningful stimulation across participants, those assigned to the Heatables condition underwent an individualized adjustment of NIR-IR intensity prior to the experimental session, accounting for known interindividual differences in thermal sensitivity \cite{ito_assessment_2000}.

\subsection{Experimental Design}
We employed a 2×2 mixed factorial design with one between-subjects factor (Stimulation Condition: Heatables vs. Placebo) and one within-subjects factor (Device Exposure: Device vs. No Device). To account for order effects and perceptual expectancy, exposure order (Device First vs. Control First) was counterbalanced, and half the participants received a visually identical but non-thermal placebo.

Participants were randomly assigned to one of four groups and completed two sessions: one with their assigned device and one control session without any device.

Table~\ref{tab:participant-groups} summarizes group allocation.

\begin{table}[h]
  \centering
  \caption{Participant allocation across counterbalanced experimental groups.}
  \label{tab:participant-groups}
  \begin{tabular}{lll}
    \toprule
    \textbf{Group} & \textbf{Condition Order} & \textbf{Gender (m/f)} \\
    \midrule
    A & Heatables $\rightarrow$ No Device & 3 / 3 \\
    B & No Device $\rightarrow$ Heatables & 3 / 3 \\
    C & Placebo $\rightarrow$ No Device   & 5 / 1 \\
    D & No Device $\rightarrow$ Placebo   & 4 / 2 \\
    \bottomrule
  \end{tabular}
\end{table}

\begin{figure*}[!t]
  \centering
  \includegraphics[width=\linewidth]{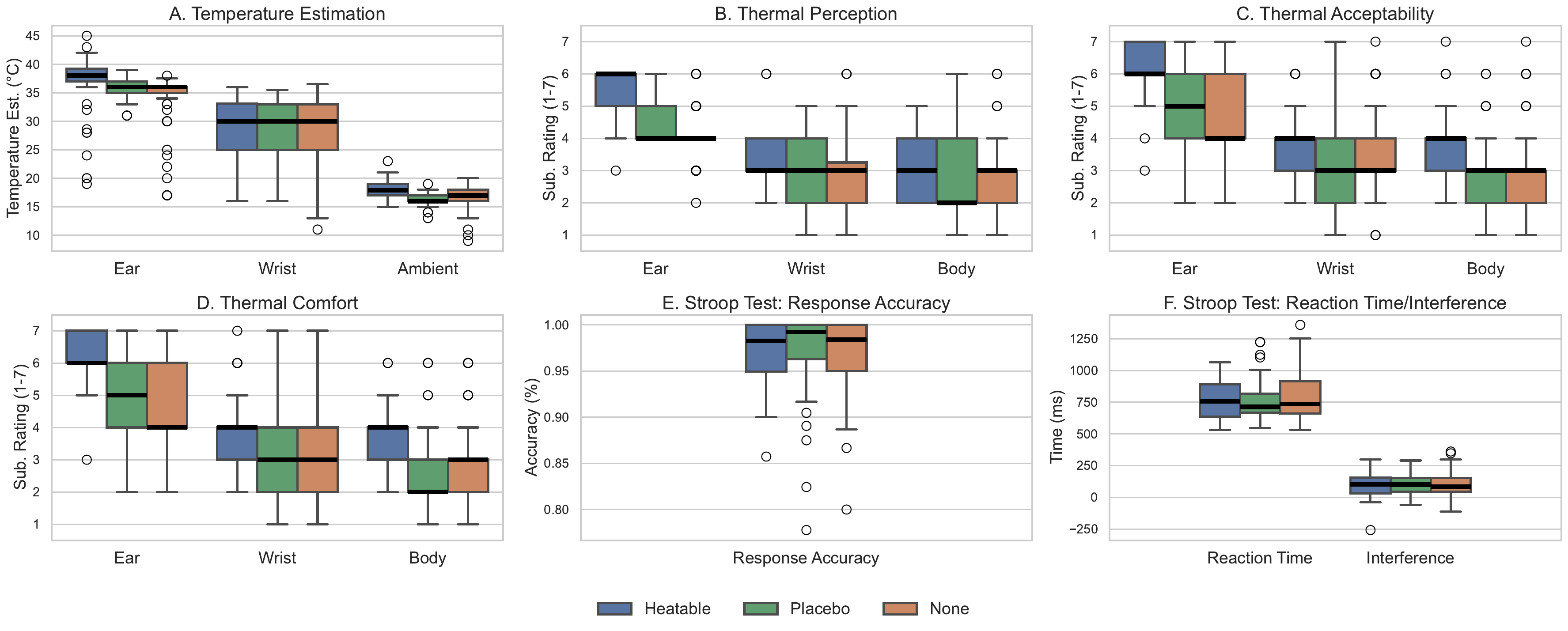}
  \caption{Box-Whisker plots of: A. temperature estimations; B.-D. subjective assessments of thermal perception, thermal acceptability, and thermal comfort; E+F. Stroop Test metrics (response accuracy and reaction time/interference). All results are presented separated for Heatable, Placebo, and control condition measurements.}
  \Description{This figure consists of six box-whisker plots, each representing outcome measures from three conditions: Heatable (blue), Placebo (green), and None/Control (orange). Each panel compares data across these groups. \begin{itemize}
      \item Panel A: Temperature Estimation
Shows estimated temperatures (in \textdegree C ) at the ear, wrist, and ambient environment.
\item Panel B: Thermal Perception (Subjective Ratings)
Displays self-reported thermal perception scores (1 = cold, 7 = hot) for ear, wrist, and body.
\item Panel C: Thermal Acceptability (Subjective Ratings)
Presents ratings of how acceptable the temperature was at each site.
\item Panel D: Thermal Comfort (Subjective Ratings)
Indicates how thermally comfortable participants felt.
\item Panel E: Stroop Test – Response Accuracy
Shows percentage accuracy the test.
\item Panel F: Stroop Test – Reaction Time and Interference (ms)
Combines reaction time and interference scores.
  \end{itemize}}
  \label{fig:box_plots}
\end{figure*}

\begin{figure*}[!t]
  \centering
  \includegraphics[width=\linewidth]{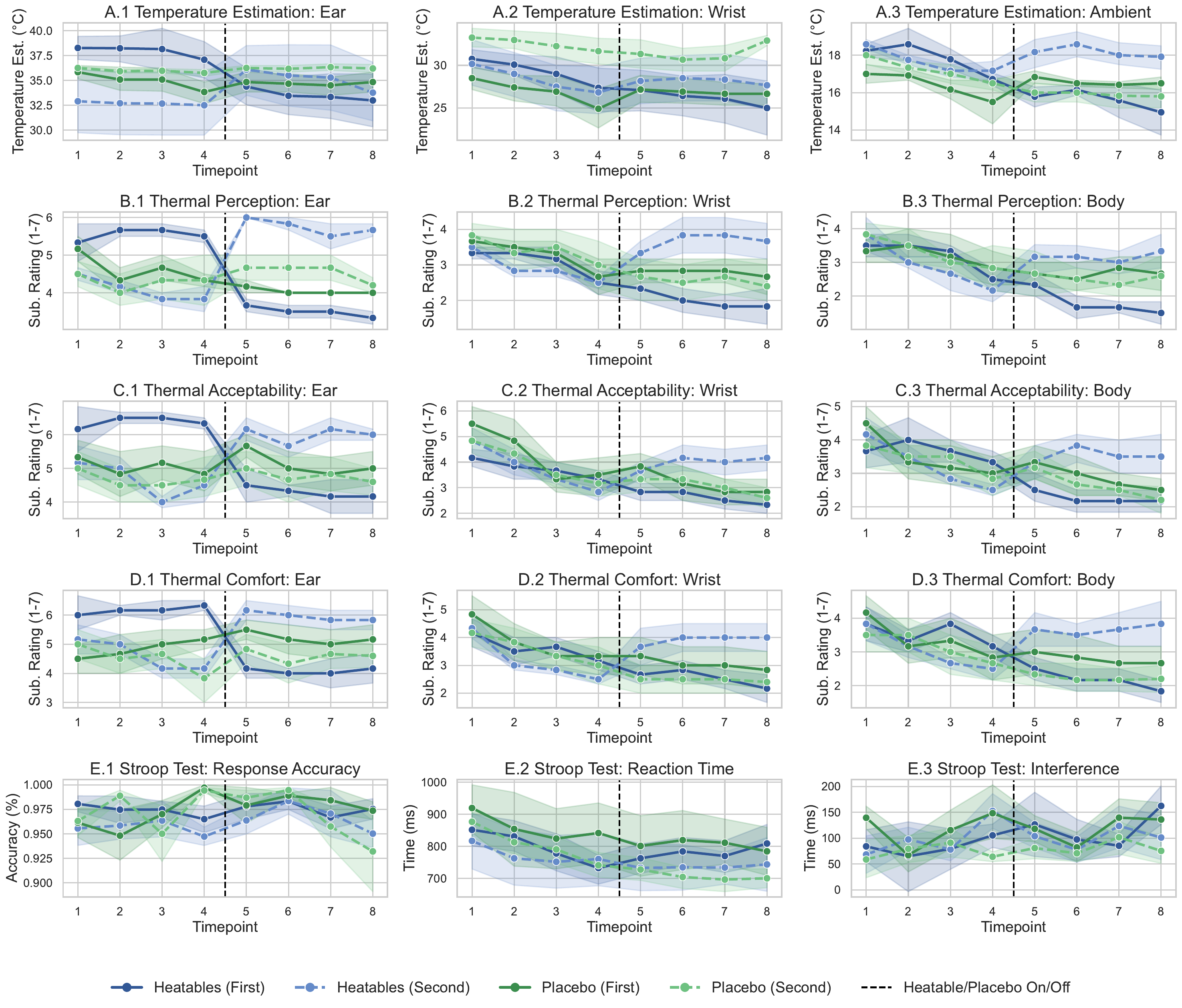}
  \caption{Temporal progression of: A. temperature estimations; subjective assessements of B. temperature perception, C. temperature acceptability, and D. temperature comfort; E. Stroop Test metrics. Results are presented across eight timepoints, separated by device (Heatables, Placebo) and exposure order. The shaded areas indicate 68\% confidence intervals. The dashed line marks the point of device attachment or removal.}
  \Description{The figure is a multi-panel line graph composed of five rows (labeled A through E) and three columns, totaling 15 subplots. Each subplot displays data across eight time points labeled “Timepoint” on the x-axis. Each row represents a different measured domain:
\begin{itemize}
    \item Row A shows objective temperature estimations (in \textdegree C ) at the ear (A.1), wrist (A.2), and ambient environment (A.3).
    \item Row B presents subjective thermal perception ratings (scale 1–7) at the ear (B.1), wrist (B.2), and whole body (B.3).
    \item Row C shows subjective thermal acceptability ratings (1–7 scale) at the same three locations (ear, wrist, body).
    \item Row D illustrates subjective thermal comfort ratings (also on a 1–7 scale) at the same three locations.
    \item Row E reports Stroop Test performance metrics: response accuracy (E.1, percentage), reaction time (E.2, milliseconds), and interference (E.3, milliseconds).
\end{itemize}}
  \label{fig:line_plots}
\end{figure*}

\subsection{Study Procedure}

The procedural sequence followed for each participant is illustrated in Figure~\ref{fig:study-procedure}:

\begin{enumerate}
    \item \textbf{Baseline Assessment (20 min):} Participants provided informed consent, demographic data, and  thermal insulation ratings.
    \item \textbf{Setup Phase (10 min):} Participants inserted the device and underwent an individualized adjustment of NIR-IR intensity. All participants subsequently engaged in a passive 10-minute media viewing task to stabilize thermal conditions.
    \item \textbf{Stimulation Phase (45 min):} Participants engaged in self-directed but sedentary desk-based tasks resembling everyday academic work, including reading, literature searches, scientific writing, and coding. These tasks were chosen to reflect naturalistic work contexts while maintaining low physical activity. Measurements were taken at four time points (T1–T4).
    \item \textbf{Recovery Phase (20 min):} A structured rest phase allowed thermal and physiological parameters to return to baseline.
    \item \textbf{Control Phase (45 min):} Participants repeated the same task protocol without wearing any device, with measurements again collected at four time points (T5–T8).
\end{enumerate}

\subsection{Measurements and Metrics}

Objective and subjective data were collected at eight sequential timepoints (T1–T8). This consistent indexing allowed direct comparisons between exposure conditions while accounting for session timing differences.

\subsubsection{Temperature Data}

Physiological measurements included tympanic temperature (Braun ThermoScan 7), wrist surface temperature (Bosch UniversalTemp infrared thermometer), and ambient temperature (Testo 610 thermo-hygrometer).

\subsubsection{Thermal Estimation and Subjective Perception}
Cognitive performance was assessed using a computerized Stroop Color and Word Test, administered at each of the eight measurement timepoints (T1–T8). Each session was preceded by a brief training block. The test comprised 40 randomized trials per timepoint, including both congruent and incongruent conditions. Participants responded via keyboard, and each trial remained on screen until a response was made. Sessions were self-paced and typically lasted under two minutes. Across the experiment, participants completed a total of 320 Stroop trials (8 sessions × 40 trials), allowing for within-subject tracking of performance over time. From each session, we extracted response accuracy, mean reaction time (RT), and Stroop interference scores (incongruent minus congruent RT), indexing selective attention, processing speed, and cognitive control.


\subsubsection{User Experience Evaluation}
After completing the experimental session (with Heatables or Placebo), participants filled out a user experience (UX) questionnaire assessing device comfort, usability, perceived safety, thermal effectiveness, and future usage intentions, using 7-point Likert scales based on the standardized Comfort Rating Scale (CRS) \cite{knight_tool_2005}. Open-ended questions captured qualitative feedback on ergonomics, thermal sensations, and potential application contexts, offering insights for future ergonomic and functional optimizations.

\section{Results}

We present the results of our evaluation, temperature estimations, subjective assessments, cognitive performance outcomes, and user experience feedback.

\subsection{Sample}
\label{sec:sample}
The final sample consisted of $N = 24$ participants (9 female, 15 male), with a mean age of $M = 28.33$ $(SD = 5.59)$.

To verify that ambient temperatures were consistent across experimental conditions, we compared the measured temperatures for each condition using a one-way ANOVA. No significant differences were found across conditions, $F(3, 196) = 1.16$ $(p = 0.338)$, indicating comparable baseline temperatures.


\subsection{Statistical Model}

For all statistical analyses but the UX reports (which are a purely descriptive measure), we employed linear mixed-effects models (LMMs). The fixed effects in each model included binary variables indicating whether the \textit{Heatable} was active or whether a \textit{Placebo} version was used. We furthermore included the variable \textit{Time} to capture within-session shifts in estimation/performance. Additionally, we included the binary variable \textit{Order}, indicating whether participants experienced the treatment (Heatable/Placebo) or control condition first.

We also incorporated random intercepts for each \textit{Participant} and random slopes for \textit{Time} in the models. This enabled the modeling of both the heterogeneity in participants' initial responses as well as their potential variation in time-related trends.

Prior to model estimation, we evaluated the assumptions of normality of residuals and homogeneity of variance. As these assumptions were 
not adequately met, all model estimates and standard errors were obtained via nonparametric bootstrapping with 1{,}000 iterations to ensure robustness against violations of normality and heteroskedasticity assumptions. 

As we have formulated directed hypotheses, we applied unidirectional testing \cite{churchill_marketing_2002, pfaffenberger_statistical_1987, jones_test_1952, cho_is_2013}. Since these hypotheses are exclusively formulated for the device effects, we will only report the results for \textit{Heatable} and \textit{Placebo} but leave out the effects of \textit{Time} and \textit{Order} to maintain conciseness. To account for the testing of each indicator at/for three separate locations/measures, the significance level $\alpha$ was adjusted to 0.017 using the Bonferroni correction.

The results from the LMMs are visualized in Figure \ref{fig:box_plots} and Figure ~\ref{fig:line_plots}.

\subsection{Temperature Estimation}

All temperature estimates were reported in Celsius. Throughout all locations, the \textit{Heatable} led to significantly higher estimates in temperature ($b_{ear} = 3.42$, $p < 0.001$; $b_{wrist} = 1.23$, $p = 0.010$; $b_{ambient} = 1.20$, $p < 0.001$). This effect was not found in the \textit{Placebo} condition, where the slopes were lower throughout and not statistically significant on the chosen significance level ($b_{ear} = 0.83$, $p = 0.020$; $b_{wrist} = -0.45$, $p = 0.508$; $b_{ambient} = 0.14$, $p = 0.375$).

\subsection{Subjective Assessments}

In the following, we report the results for the three measures taken in the survey. All assessments were given on a 7-point Likert scale.

\subsubsection{Thermal Perception}
The perception of temperature was significantly higher for all locations when the \textit{Heatable} was worn ($b_{ear} = 1.97$, $p < 0.001$; $b_{wrist} = 0.66$, $p = 0.013$; $b_{body} = 0.71$, $p = 0.007$). For the \textit{Placebo}, this effect was only observed for perception of ear temperature ($b_{ear} = 0.43$, $p = 0.015$; $b_{wrist} = 0.13$, $p = 0.280$; $b_{body} = 0.41$, $p = 0.049$).

\subsubsection{Thermal Acceptance}
The acceptance of temperature was significantly higher for both the ear and the body when the \textit{Heatable} was worn, but not for the wrist ($b_{ear} = 1.78$, $p < 0.001$; $b_{wrist} = 0.68$, $p = 0.019$; $b_{body} = 1.04$, $p < 0.001$). Again, the \textit{Placebo} only had this effect for the assessment at the ear ($b_{ear} = 0.58$, $p = 0.006$; $b_{wrist} = 0.45$, $p = 0.032$; $b_{body} = 0.31$, $p = 0.130$).

\subsubsection{Thermal Comfort}
The reported comfort was significantly higher for all locations when the \textit{Heatable} was attached ($b_{ear} = 1.95$, $p < 0.001$; $b_{wrist} = 0.83$, $p = 0.016$; $b_{body} = 0.99$, $p < 0.001$). For the \textit{Placebo}, the effect on comfort was only observed at the ear ($b_{ear} = 0.42$, $p = 0.006$; $b_{wrist} = -0.01$, $p = 0.967$; $b_{body} = 0.07$, $p = 0.404$).


\subsection{Cognitive Performance}

The stroop test performances were reported in percentages for response accuracy and in milliseconds for reaction time and interference. Statistically significant effects were neither found for the \textit{Heatable} ($b_{acc} = 0.007$, $p = 0.450$; $b_{react} = -15.52$, $p = 0.218$; $b_{interf} = -12.63$, $p = 0.226$) nor the \textit{Placebo} ($b_{acc} = -0.0040$, $p = 0.618$; $b_{react} = -11.25$, $p = 0.159$; $b_{interf} = -2.09$, $p =0.437$) throughout all measured performance indicators.


\subsection{UX Reports}

To assess how participants experienced wearing the Heatables in our experimental setting, we gathered both quantitative and qualitative feedback.

\subsubsection{Quantitative Reports}
Participants generally rated the device as both comfortable ($M = 6.25$, $SD = 0.62$) and safe to use ($M = 6.50$, $SD = 0.67$), with even the lowest ratings remaining on the more favorable end of the scale. Many users reported perceiving a noticeable difference compared to regular earbuds ($M = 4.67$, $SD = 1.72$); however, this did not translate into discomfort or pain ($M = 1.25$, $SD = 0.45$). While nearly all participants indicated a willingness to use the device in private settings ($M = 5.92$, $SD = 1.51$), responses were more reserved regarding its use in public ($M = 5.33$, $SD = 1.56$). 

\subsubsection{Qualitative Reports}
Participants provided open-ended feedback reflecting their subjective experience with the device. Responses were clustered into three primary themes: comfort and sensation, perceived use cases, and physical discomfort.

\paragraph{Thermal Comfort and Sensory Experience.} Overall, the device was described as “comfortable to wear,” “warm and cozy,” and “relaxing.” Several users noted that the warming sensation extended beyond the ears to the surrounding head region, suggesting a broader thermal perception than expected. One participant remarked, “the warmth felt like the ambient temperature became warmer slightly.”

\paragraph{Perceived Use Cases.} Many participants described potential future use of the device in cold environments, such as during winter walks, cycling, or visits to outdoor markets (e.g., “bike riding in winter,” “Christmas markets,” or “under a hat during a cold winter day”). Others mentioned relaxation-oriented contexts including “studying,” “meditation,” or “illness recovery,” indicating interest in the device’s warming functionality beyond physical comfort alone.

\paragraph{Discomfort and Adaptation.} Reports of discomfort were minimal. One participant described a slight sensitivity in one ear due to the device’s bulk, while another noted that “the warmth builds up and eventually makes you want to ventilate the ears.” Some described the sensation as unfamiliar but not painful—“you have to get used to it, but it’s not painful.”

In summary, the qualitative feedback complements the quantitative UX results: participants generally perceived the device as comfortable and suitable for cold-weather scenarios, with only minor concerns related to heat buildup or physical form factor.

\section{Discussion}

\paragraph{Interpretation of Results.}
Heatables significantly enhanced subjective thermal experience: participants reported higher perceived temperatures and improved thermal comfort and acceptability. These effects extended beyond the ear to the wrist and body, despite stimulation being confined to the auditory canal. This supports the hypothesis that the ear can act as a perceptual gateway with systemic thermal influence.

Cognitive performance remained stable: Stroop accuracy was consistently high across all conditions, in line with prior work suggesting that mild cold exposure in low-demand contexts does not reliably impair cognition. The absence of performance decline supports the use of Heatables without cognitive side effects.

Participants rated the device as safe, comfortable, and unobtrusive. Most expressed willingness to use it in private or relaxation-oriented contexts (e.g., studying, recovery). Some reported mild heat buildup or a need to adapt to the sensation, but no serious discomfort. Several associated the warmth with calming or soothing effects, indicating potential applications beyond thermal comfort, such as stress modulation.

These findings demonstrate the feasibility of in-ear IR/NIR-LED-based thermal stimulation for perceptually enhancing whole-body warmth in everyday settings—without compromising usability or cognition.

\paragraph{Design Implications and Future Work.}
Heatables highlight the potential of thermal feedback as a novel modality in hearables. Future systems could combine environmental or physiological sensing with adaptive in-ear heating for personalized comfort, distraction mitigation, or mood regulation. Acceptance in public settings may improve with further miniaturization.

\paragraph{Limitations.}
While the Stroop task is well-established, it may lack sensitivity to subtle thermal effects. A ceiling effect is also possible, as accuracy scores remained consistently high. The evaluation was limited to one stable indoor condition (17.5°C), which may constrain generalizability. Moreover, the study design does not isolate radiative from conductive heating. Future work should include IR-blocking controls to disentangle optical versus passive heat effects.

\section{Conclusion} We presented \textit{Heatables}, an in-ear wearable that delivers targeted thermal stimulation using IR-NIR-LEDs. In a placebo-controlled study under cool stress (\(\approx 17.5 ^\circ\mathrm{C}\)), 
Heatables significantly increased subjective ambient and wrist temperature estimates by approximately 1.2°C (95\% CI [0.7, 1.7]) and 1.2°C (95\% CI [0.3, 2.1]), respectively, compared to placebo and control conditions. These effects suggest that localized in-ear heating can induce a broader perceptual shift beyond the stimulation site. Participants also reported a delayed onset of cold discomfort, while cognitive performance, assessed via Stroop tasks, remained unaffected. These findings demonstrate the auditory canal as a promising interface for unobtrusive and perceptually effective thermal stimulation. Heatables offer a new pathway towards highly portable, personal comfort solutions in both stationary and mobile contexts.


\begin{acks}
This work was partially funded by the ZEISS Innovation Hub and the Deutsche Forschungsgemeinschaft (DFG, German Research Foundation) – GRK 2739/1 – Project No. 447089431 – Research Training Group: KD2School – Designing Adaptive Systems for Economic Decisions.
\end{acks}


\balance
\bibliographystyle{ACM-Reference-Format}
\bibliography{bibliography}

\appendix

\end{document}